\documentclass{article}
\usepackage{spconf,amsmath,graphicx,textcomp}
\usepackage{color,amsmath,epsfig}

\def\x{{\mathbf x}}

\title{SPARSITY-AWARE FILTERED-X AFFINE PROJECTION ALGORITHMS FOR ACTIVE NOISE CONTROL}

\name{Amelia Gully {$^\diamondsuit$}  and Rodrigo C. de Lamare
{$^\diamondsuit$ $^\spadesuit$}
\thanks{Thanks to the Department of Electronics of the University of
York, and the York-FAPESP project on sparsity-aware adaptive algorithms and applications, for partially funding this work.}}
\address{Department of Electronics, University of York, UK $^\diamondsuit$   \\
CETUC, PUC-Rio, Brazil  $^\spadesuit$ \\
Emails: ajg540@york.ac.uk, rcdl500@york.ac.uk}

\begin{document}

\maketitle

\begin{abstract}

This paper describes a novel technique for promoting sparsity in the
modified filtered-x algorithms required for active noise control.
The proposed algorithms are based on recent techniques incorporating
approximations to the \(\ell_0\)-norm in the cost functions that are
used to derive adaptive filtering algorithms. In particular,
zero-attracting and reweighted zero-attracting filtered-x adaptive
algorithms are developed and considered for active noise control
problems. The results of simulations indicate that the proposed techniques improve
the convergence of the existing modified algorithm in the case where
the primary and secondary paths exhibit a degree of sparsity.
\end{abstract}
\section{INTRODUCTION}
\label{sec:intro}

Active noise control (ANC) is a popular technique for removing noise
from a system by  using a variant of the system identification
scenario to subtract the effect of a noise-generating plant from a
signal \cite{ancPrinc}. ANC algorithms are based on classical
adaptive algorithms, with provision made for the additional
electroacoustic path between the filter output and measured error signal,
known as the secondary path \cite{ancTutRev}. A common technique to
account for this path is known as the filtered-x (FX) scheme,
originally developed for least-mean square (LMS) algorithms in the
context of adaptive control \cite{adInvCont} and since developed for
other classical algorithms \cite{adSigProc,fapAnc,ancSystems}.

The FX scheme eliminates potential algorithm instability caused by
the additional delay in the secondary path \cite{analysisFxLMS}, but
is marred by slow convergence \cite{improvedMFxLMS}. The modified
filtered-x (MFX) scheme was introduced in \cite{modFxLMS}, and
improves upon the convergence of the FX algorithms by introducing
additional steps to approximate the error signal. MFX schemes have
since become a popular choice for ANC architectures. However, there
is still room to improve the convergence speed.

Recent developments in the field of compressive sensing have led to
the introduction of sparsity-promoting penalties in adaptive
filtering algorithms
\cite{sparseLMS,sparseL0lms,sparseAP,delamaresp,jio,jidf,jio_stcdma,sa_stap},
producing zero-attracting (ZA) and reweighted zero-attracting (RZA)
algorithms. These have been shown to provide faster convergence when
the system in question has a degree of sparsity. This is often the
case in the real systems encountered in ANC applications. To the
knowledge of the authors, however, this powerful sparsity-inducing
technique has not been considered with the modified filtered-x
architecture used in ANC.

In this paper, sparsity-inducing techniques are incorporated
into the filtered-x affine projection (FxAP) algorithm. The FxAP
algorithm is selected since it is a popular algorithm for ANC
applications, performing well in the presence of correlated signals,
and allowing a trade-off between convergence speed and complexity to
be easily controlled by varying the projection order \(K\). A
number of techniques have been introduced to improve the FxAP
algorithm, including the introduction of the modified form, MFxAP
\cite{fapAnc}, and adapting the system parameters over time
(see \cite{evolvingFxAP} and references therein). This paper proposes
zero-attracting and reweighted zero-attracting MFxAP algorithms.
Experimental results demonstrate the superior performance of the
proposed algorithms for systems with a moderate degree of sparsity.

The rest of this paper is structured as follows. In Section 2, the
active noise control problem is stated and the MFxAP algorithm is
briefly reviewed. In Section 3 the proposed algorithms are developed
with reference to the MFxAP algorithm from which they derive. In
Section 4, the results of simulation trials performed on these
algorithms are presented and discussed. Section 5 presents
conclusions and describes avenues of further research.

\textit{Notation}: Throughout this paper, uppercase boldface letters
will be used to denote matrices, and lowercase boldface letters to
denote vectors. \(E\lbrace\cdot\rbrace\) denotes statistical
expectation, \(\text{sgn}\lbrace\cdot\rbrace\) is the signum
function, \(\mathbf{I}\) is an identity matrix of appropriate
dimensions, and \(\lVert\cdot\rVert_p\) denotes the \(p\text{th}\)
norm of a vector. All vectors are column vectors. Following
convention, the term \(\ell_0\)-norm and notation
\(\lVert\cdot\rVert_0\) denotes the number of non-zero elements in a
vector.

\section{Problem Statement and Review of the MFxAP Algorithm}
\label{sec:algs}

In this section, the problem of active noise control is explained
and the MFxAP algorithm is reviewed.

\begin{figure}[t]
\begin{center}
\def\epsfsize#1#2{1\columnwidth}
\epsfbox{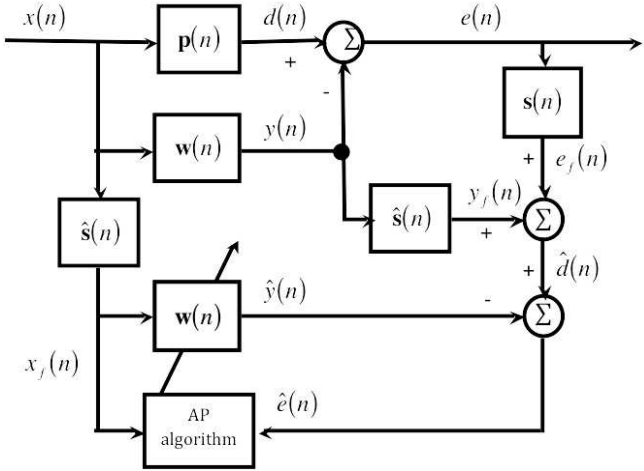} \caption{Modified filtered-x AP structure.}
\end{center}
\end{figure}

In the common case of feedforward ANC, an error signal is used to adjust an adaptive filter such that its output can be `subtracted' from a signal using the principle of destructive interference, and hence eliminate or significantly reduce disturbing noise. However, the electroacoustic nature of ANC systems introduces a combined \textit{secondary path}, since each component has an associated transfer function. It was shown in \cite{analysisFxLMS} that this can be split into two parts, one estimated as part of the plant, and one occuring after the summing junction and affecting the error signal, here denoted \(\mathbf{s}(n)\). The result is a delay in the system that may cause conventional adaptive algorithms to become unstable. The FX algorithms account for this by using an estimate of the secondary path, \(\mathbf{\hat{s}}(n)\), to filter the input signal \(x(n)\), so that the algorithm input is the filtered signal \(x_f(n)\). However, FX algorithms have strict limits on step size that negatively impact upon convergence speed \cite{improvedMFxLMS}. The MFX algorithms improve upon this convergence speed by estimating the actual error signal, but can be improved further by introducing variable parameter or sparsity-inducing techniques such as those proposed here. The derivation of the MFxAP algorithm is reviewed below.

When deriving the conventional AP algorithm, the following
expression is minimised:
\[\lVert\mathbf{w}(n+1)-\mathbf{w}(n)\rVert^2\quad\text{s.t.}\quad\mathbf{d}(n)=\mathbf{U}(n)\mathbf{w}(n+1)\]where
\(\mathbf{U}(n)\) is a regressor matrix containing past values of
\(x(n)\) as follows:
\[\setlength{\arraycolsep}{2pt}\renewcommand{\arraystretch}{0.8}
\mathbf{U}(n)=\begin{bmatrix}x(n)& x(n-1)& \cdots& x(n-L+1)\\x(n-1)&
x(n-2)& \cdots& x(n-L)\\\vdots& \vdots& \ddots& \vdots\\x(n-K+1)&
x(n-K)& \cdots& x(n-L-K+2)\end{bmatrix}\]
The MFxAP algorithm is
well suited to ANC as it does not require that \(\mathbf{d}(n)\) be
available for its derivation. Instead, we set the condition to be
\begin{equation}\label{condition1}
\mathbf{\hat{d}}(n)=\mathbf{U}_f(n)\mathbf{w}(n+1)
\end{equation}
where, as seen in Fig. 1,
\(\mathbf{\hat{e}}(n)=\mathbf{\hat{d}}(n)-\mathbf{\hat{y}}(n)\).
This uses a \textit{filtered} regressor matrix, \(\mathbf{U}_f(n)\),
containing past values of \(x_f(n)\). Therefore, the condition
states that the \textit{estimated} output for the next iteration
should equal the \textit{estimated} desired signal for the current
iteration. Then, using a vector of Lagrange multipliers,
\(\underline{\lambda}\), the cost function becomes:
\begin{equation}\label{jmfxap}
\begin{split}
J_{1}(n)=&\lVert\mathbf{w}(n+1)-\mathbf{w}(n)\rVert^2\\&+\text{Re}\lbrace[\mathbf{\hat{d}}(n)-\mathbf{U}_f(n)\mathbf{w}(n+1)]^H\underline{\lambda}\rbrace
\end{split}
\end{equation}
In order to minimise the above cost function, we equate the gradient of \(J_{1}(n)\)
with respect to \(\mathbf{w}^*(n+1)\) to zero:
\[\frac{\partial J_{1}(n)}{\partial\mathbf{w}^*(n+1)}=\mathbf{w}(n+1)-\mathbf{w}(n)-\mathbf{U}_f^H(n)\underline{\lambda}=0\]
\begin{equation}\label{mfxapw}
\mathbf{w}(n+1)=\mathbf{w}(n)+\mathbf{U}_f^H(n)\underline{\lambda}
\end{equation}
Substituting the above expression into \eqref{condition1}, gives:
\[\mathbf{\hat{d}}(n)=\mathbf{U}_f(n)\mathbf{w}(n)+\mathbf{U}_f(n)\mathbf{U}_f^H(n)\underline{\lambda}\]
\[\mathbf{\hat{d}}(n)=\mathbf{\hat{y}}(n)+\mathbf{U}_f(n)\mathbf{U}_f^H(n)\underline{\lambda}\]
\[\mathbf{\hat{e}}(n)=\mathbf{U}_f(n)\mathbf{U}_f^H(n)\underline{\lambda}\]
\begin{equation}\label{mfxaplam}
\underline{\lambda}=[\mathbf{U}_f(n)\mathbf{U}_f^H(n)]^{-1}\mathbf{\hat{e}}(n)
\end{equation}
Inserting \eqref{mfxaplam} back into \eqref{mfxapw}, and incorporating a step-size parameter, \(\mu\), and a small positive constant, \(\delta\), to avoid inverting a singular matrix, the MFxAP recursion is obtained:
\begin{equation}\label{mfxap}
\mathbf{w}(n+1)=\mathbf{w}(n)+\mu\mathbf{U}_f^H(n)[\mathbf{U}_f(n)\mathbf{U}_f^H(n)+\delta\mathbf{I}]^{-1}\mathbf{\hat{e}}(n)
\end{equation}

\section{Proposed Sparsity-Inducing Filtered-X Affine Projection
Algorithms}

In this section, the proposed sparsity-inducing filtered-x affine
projection algorithms are derived. In particular, the
zero-attracting and reweighted zero-attracting strategies
considered in \cite{sparseLMS,sparseAP} are incorporated into a
filtered-x structure, resulting in new algorithms for active noise
control.

\subsection{Zero-attracting MFxAP (ZA-MFxAP) algorithm}
\label{ssec:zafxap}

The ZA-MFxAP incorporates a sparsity-inducing penalty to attract coefficients towards zero. The obvious choice for such a penalty would be the \(\ell_0\)-norm. However, this is a discontinuous and non-convex function, making it costly and difficult to optimise mathematically. Zero-attracting algorithms use an \(\ell_1\)-norm penalty as an approximation, and the penalty is added to the cost function in \eqref{jmfxap} as follows:
\begin{equation}\label{jzamfxap}
\begin{split}
J_{2}(n)=&\lVert\mathbf{w}(n+1)-\mathbf{w}(n)\rVert^2\\
&+\text{Re}\lbrace[\mathbf{\hat{d}}(n)-\mathbf{U}_f(n)\mathbf{w}(n+1)]^H\underline{\lambda}\rbrace\\
&+\alpha\lVert\mathbf{w}(n+1)\rVert_1
\end{split}
\end{equation}
As before, this cost function is minimised with respect to \(\mathbf{w}^*(n+1)\) and equated to zero, giving:
\[\mathbf{w}(n+1)=\mathbf{w}(n)+\mathbf{U}_f^H(n)\underline{\lambda}-\alpha\text{sgn}\lbrace\mathbf{w}(n+1)\rbrace\]
Solving for \(\underline{\lambda}\) and incorporating \(\mu\) and \(\delta\) as before, and making the assumption that \(\text{sgn}\lbrace\mathbf{w}(n+1)\rbrace\approx\text{sgn}\lbrace\mathbf{w}(n)\rbrace\), we obtain the ZA-MFxAP recursion:
\begin{equation}\label{zamfxap}
\begin{split}
\mathbf{w}(n+1)=&\mathbf{w}(n)+\mu\mathbf{U}_f^+(n)\mathbf{\hat{e}}(n)\\&+\rho\mathbf{U}_f^+(n)\mathbf{U}_f(n)\text{sgn}\lbrace\mathbf{w}(n)\rbrace\\&-\alpha\text{sgn}\lbrace\mathbf{w}(n)\rbrace
\end{split}
\end{equation}
where \(\rho=\mu\alpha\) is known as the zero-attraction strength,
and
\(\mathbf{U}_f^+(n)=\mathbf{U}_f^H(n)[\mathbf{U}_f(n)\mathbf{U}_f^H(n)+\delta\mathbf{I}]^{-1}\).
This is identical to \eqref{mfxap} with the addition of two terms,
controlled by \(\rho\), which attract the coefficients towards zero.

\subsection{Reweighted zero-attracting MFxAP (RZA-MFxAP) algorithm}
\label{ssec:rzafxap}

The RZA-MFxAP algorithm uses a log-sum penalty in place of the \(\ell_1\)-norm, as this provides a closer approximation to the behaviour of the \(\ell_0\)-norm. Thus, the cost function becomes:
\begin{equation}\label{jrzamfxap}
\begin{split}
J_{3}(n)=&\lVert\mathbf{w}(n+1)-\mathbf{w}(n)\rVert^2\\
&+\text{Re}\lbrace[\mathbf{\hat{d}}(n)-\mathbf{U}_f(n)\mathbf{w}(n+1)]^H\underline{\lambda}\rbrace\\
&+\gamma\sum_{i=1}^L \log(1+\epsilon\lvert w_i(n) \rvert)
\end{split}
\end{equation}
Following a similar derivation procedureto ZA-MFxAP, the RZA-MFxAP recursion is given by:
\begin{equation}\label{rzamfxap}
\begin{split}
\mathbf{w}(n+1)=&\mathbf{w}(n)+\mu\mathbf{U}_f^+(n)\mathbf{\hat{e}}(n)\\&+\rho'\mathbf{U}_f^+(n)\mathbf{U}_f(n)\mathbf{\Psi}(n)-\gamma\epsilon\mathbf{\Psi}(n)
\end{split}
\end{equation}
where \(\mathbf{\Psi}(n) =
\frac{\text{sgn}\lbrace\mathbf{w}(n)\rbrace}{1+\epsilon\lvert\mathbf{w}(n)\rvert}\).

In this case the strength of the zero-attraction is controlled by
\(\rho'=\mu\gamma\epsilon\) where \(\epsilon\) is known as the
shrinkage magnitude. The RZA-MFxAP is more selective than the
ZA-MFxAP, and attracts coefficients proportional to
\(^1/_\epsilon\) toward zero most strongly.
Therefore careful tuning of \(\epsilon\) can reduce the bias of the
estimation procedure.

\begin{figure}[t]
\begin{center}
\def\epsfsize#1#2{1\columnwidth}
\epsfbox{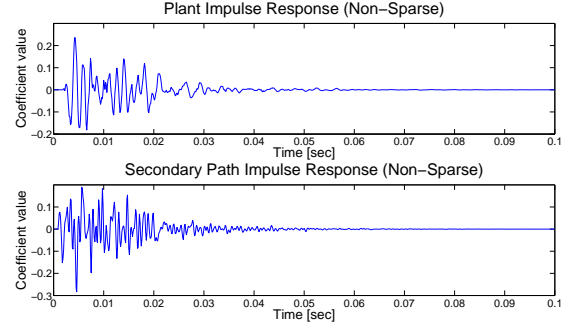}  \caption{Non-sparse plant and secondary paths.}
\end{center}
\end{figure}

\section{RESULTS}
\label{sec:results}

This section compares the results of simulation trials for the
proposed algorithms outlined in \eqref{zamfxap} and \eqref{rzamfxap} with those
of the conventional FxAP and MFxAP algorithms. The results are
averaged over 50 simulation trials.

Three types of primary path were used: a non-sparse path (density 785/800), illustrated in Fig. 2, a partially-sparse path formed by arbitrarily setting the majority of the coefficients in the non-sparse path to zero (density 73/800) and a sparse path in which the fourth coefficient is set to one and all remaining coefficients to zero (density 1/800). Three types of secondary path with similar densities were generated in the same way from the non-sparse secondary path in Fig. 2, with \(\mathbf{\hat{s}}(n)=\mathbf{s}(n)\). A poorer estimate of  \(\mathbf{s}(n)\) is known to degrade convergence speed, although it has been shown in \cite{imperfectSecPath} that in rare cases \(\mathbf{\hat{s}}(n)\neq\mathbf{s}(n)\) may improve algorithm performance.

 For each primary path formulation, the algorithms were run for 220,000 iterations with the secondary path set as sparse at the start of the experiment, changed to partially-sparse at iteration 10,000 and to non-sparse at iteration 70,000. These values were found to provide sufficient time for the proposed algorithms to converge. Figures 3-5 show mean-square deviation (MSD) convergence curves for each of these experiments. Each secondary path change requires that \(\mu\) and the parameter \(\epsilon\) in the RZA-MFxAP algorithm be retuned, with
the requisite values given in the figures. The values \(\delta =
0.002\) and \(\rho = 0.0000001\) were found to be suitable for almost all applications, with \(\rho\) requiring retuning only for the ZAMFxAP algorithm in Fig. 5.

The first 10,000 iterations in each figure illustrate algorithm performance
when the secondary path is sparse. In all cases, the convergence
speeds of conventional and proposed algorithms are similar, with the
higher-order RZA-MFxAP algorithms showing a slight improvement upon
FxAP and MFxAP when the plant is partially sparse as in Fig. 4.

\begin{figure}[t]
\begin{center}
\def\epsfsize#1#2{1\columnwidth}
\epsfbox{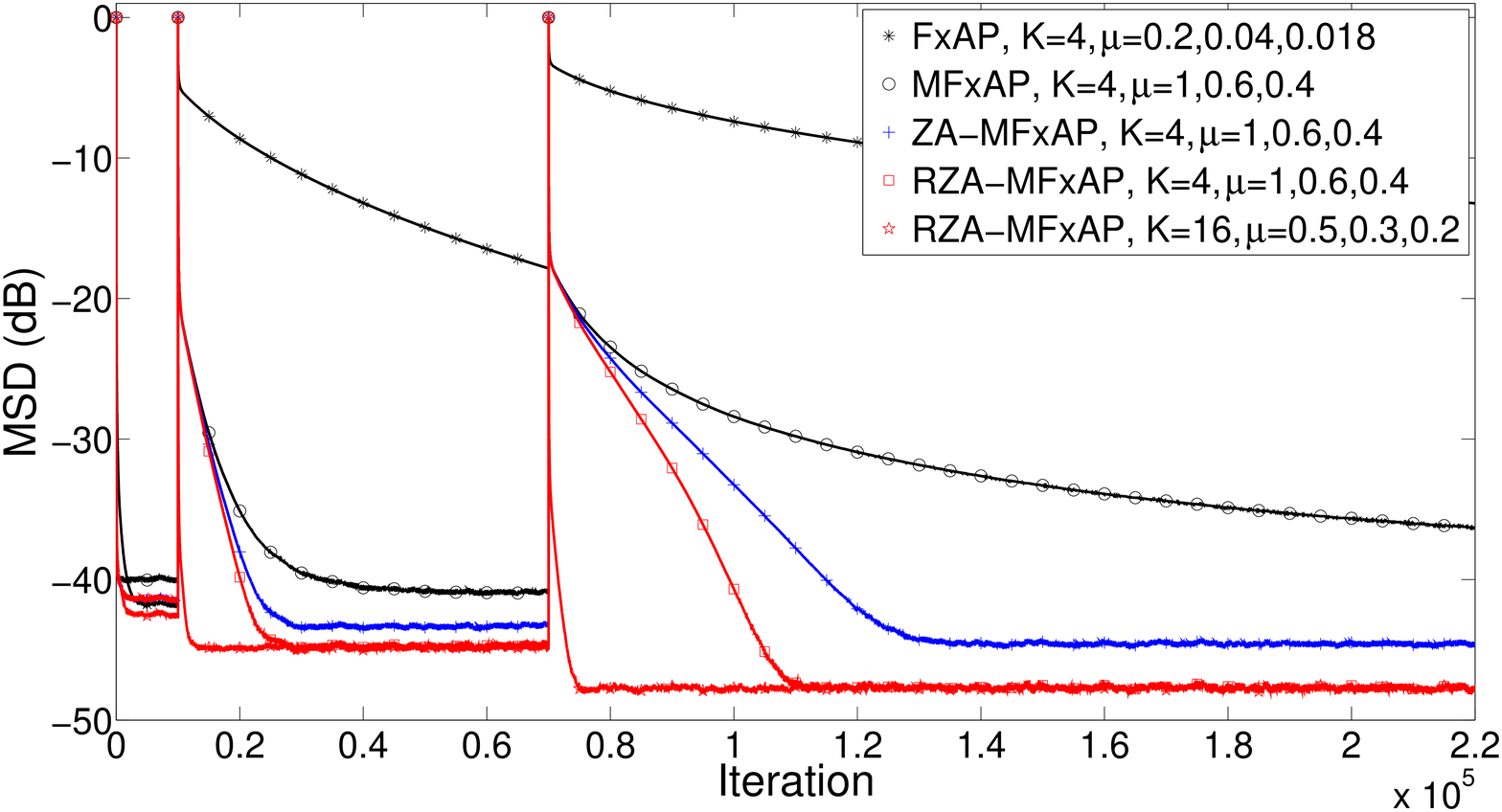}  \caption{MSD results for sparse plant
(\(\rho=0.0000001\) and \(\epsilon=10\) unless stated otherwise).}
\label{fig:SINR_RO_nom}
\end{center}
\end{figure}

\begin{figure}[t]
\begin{center}
\def\epsfsize#1#2{1\columnwidth}
 \epsfbox{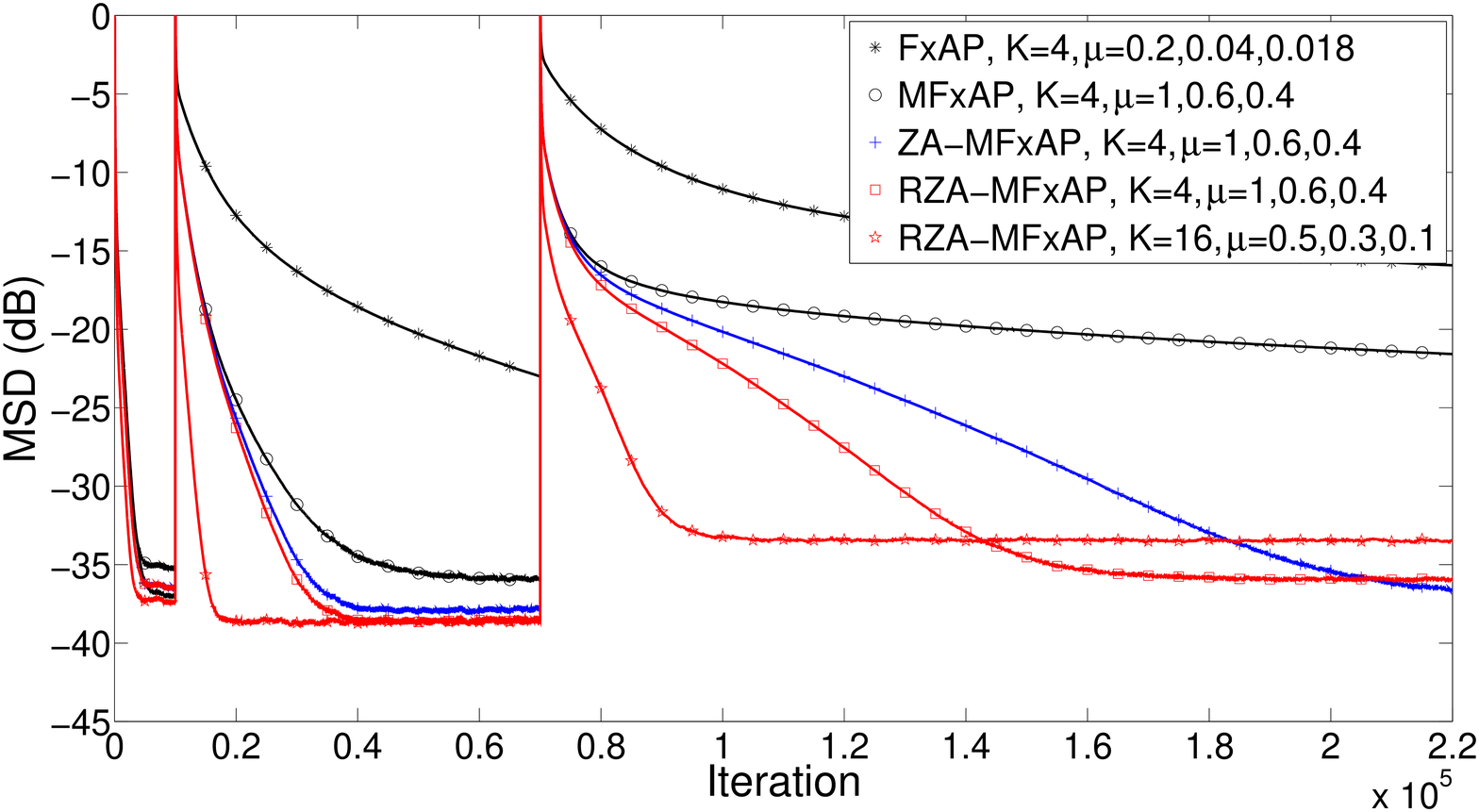}
\caption{MSD results for partially-sparse plant (\(\rho=0.0000001\)
and \(\epsilon=10\) unless stated otherwise).}
\end{center}
\end{figure}

It is seen in Fig. 3 that as the secondary path density increases, the convergence speed of the FxAP and MFxAP decreases significantly. The ZA-MFxAP and RZA-MFxAP with projection order 4 improve upon these algorithms by converging within far fewer iterations, but the higher-order RZA-MFxAP algorithm reduces convergence time to a very low number of iterations and also attains lower steady-state MSD than the other algorithms, indicating that the comparative density of the secondary path has a less significant effect on this algorithm than conventional ANC algorithms.

\begin{figure}[t]
\begin{center}
\def\epsfsize#1#2{0.95\columnwidth}
\epsfbox{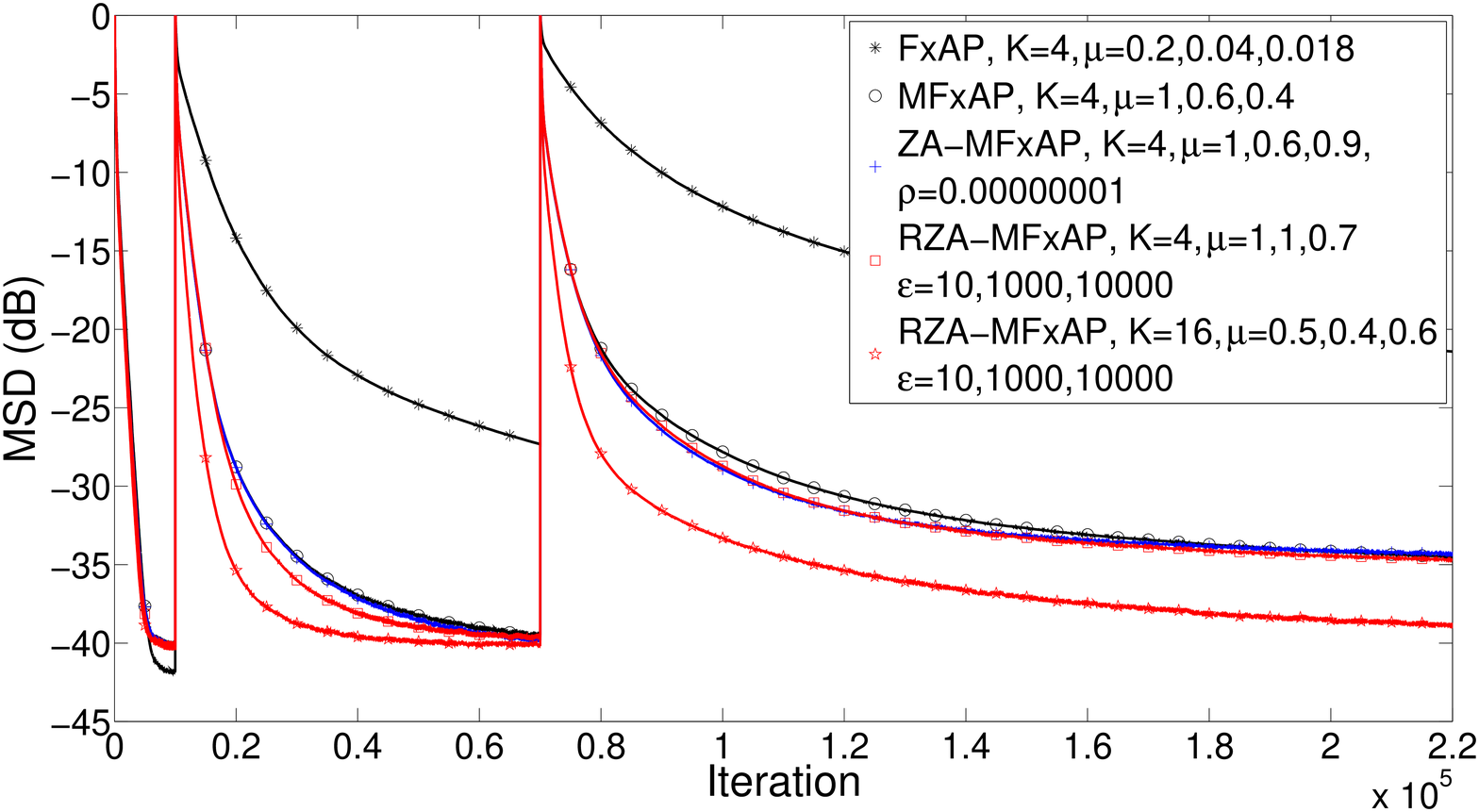}
\caption{MSD results for non-sparse plant
(\(\rho=0.0000001\) and \(\epsilon=10\) unless stated otherwise).}
\end{center}
\end{figure}

The performance of the algorithms when the plant is partially sparse
can be seen in Fig. 4. When the secondary path is semi-sparse, the proposed
algorithms show clear improvement - in terms of both MSD convergence
speed and steady-state performance - over the traditional MFxAP, and
vastly outperform the FxAP algorithm. As the secondary path density
increases further in the third section of Fig. 4, the performance of
the MFxAP is significantly degraded, whereas the proposed algorithms
continue to perform far better, with the RZA-MFxAP - particularly at
higher order - showing the fastest convergence. It should be noted that every algorithm
initially converges at the same rate, but the RZA-MFxAP algorithms
increase in convergence speed as the selective aspect begins work on
coefficients proportional to \(1\textfractionsolidus\epsilon\)
\cite{sparseLMS}. Steady-state differences are clearly visible between RZA-MFxAP orders in this section of Fig. 4. This is primarily due to the tuning of \(\epsilon\), which has a significant impact upon convergence speed if reduced from the selected value. It would therefore be worthwhile to consider algorithms incorporating a variable shrinkage magnitude parameter as an improvement to the algorithms proposed here, allowing steady-state performance to be improved without negatively affecting convergence rate.

The results of running the algorithms for a non-sparse plant can be
seen in Fig. 5. In this case, while the proposed algorithms continue
to significantly outperform the FxAP, the performance gain over the
MFxAP algorithm is less pronounced. This is because the plant is non-sparse, and therefore
the sparsity-inducing technique does not give the proposed
algorithms an advantage when finding an estimate. Although the
difference is less notable than in previous cases, the proposed
algorithms do outperform the MFxAP somewhat - particularly the
RZA-MFxAP with a projection order of 16 - due to the presence of close-to-zero coefficients in the
plant impulse response.

These results demonstrate that in any case where a degree of
sparsity is to be expected in the primary or secondary path, the
best convergence speed and steady-state error can be obtained by
using the proposed RZA-MFxAP algorithm, particularly at a relatively
high projection order such as 16. A higher order may be undesirable
from a complexity point of view, but since the algorithms have a
complexity on the order of \(L^2\) and \(K \ll L\), the increase in
complexity will be small in proportion to the performance gain.
Further work might incorporate recent improvements to sparse
techniques \cite{sa-alt}.

\section{CONCLUSION}
\label{sec:conc}

This paper has proposed adaptive algorithms for exploiting sparsity
in modified filtered-x algorithms required in active noise control
problems. The proposed algorithms have incorporated zero-attracting
and reweighted zero-attracting strategies into filtered-x adaptive
algorithms. The results of simulations have shown that the proposed
techniques improve the convergence of the existing modified
algorithm in the case where the primary and secondary paths exhibit
a degree of sparsity.

\vfill\pagebreak

\bibliographystyle{IEEEbib}
\bibliography{refs}

\end{document}